\documentclass[reprint,pre]{revtex4-2}
\usepackage{graphicx}
\usepackage{color}
\usepackage{amsmath}
\usepackage{amssymb}
\graphicspath{{figure/}}
\begin{document}

\title{Computing diffraction anomalies as nonlinear eigenvalue problems}

\author{Zitao Mai}
\author{Ya Yan Lu}
\email{mayylu@cityu.edu.hk}
\affiliation{Department of Mathematics, City University of Hong Kong, 
  Kowloon, Hong Kong, China} 
\date{\today}

\begin{abstract}
When a plane  electromagnetic wave impinges upon a diffraction grating or other
periodic structures, reflected and transmitted waves propagate away
from the structure in different radiation channels.
A diffraction anomaly occurs when
the outgoing waves in one or more radiation channels vanish. 
Zero reflection, zero transmission and perfect absorption are
important examples of diffraction anomalies, and they are useful 
for manipulating electromagnetic  waves and light.
Since diffraction anomalies appear only at specific frequencies and/or
wavevectors, and may require the tuning of structural or material 
parameters, they are relatively difficult to find by standard numerical
methods. Iterative methods may be used, but good initial
guesses are required. To determine all diffraction anomalies in a given frequency
interval, it is necessary to repeatedly solve the
diffraction problem for many frequencies. In this paper, an efficient
numerical method is developed for 
computing diffraction anomalies. The method relies on 
nonlinear eigenvalue formulations for scattering anomalies and
solves the nonlinear eigenvalue problems by a contour-integral
method. Numerical examples involving periodic arrays of cylinders 
are presented to illustrate the new method. 
\end{abstract}
\maketitle

\section{Introduction}
For diffraction  gratings and other periodic structures, transmission
and reflection spectra often have interesting and useful features,
such as sharp peaks and dips, flat bands, and rapid variation from
a peak to a dip~\cite{wood,fano41,hessel}. For structures without
material loss, especially when the structures have relevant symmetry,
it is possible to have zero reflection (total transmission) or zero 
transmission (total reflection), for some special frequency and
wavevector~\cite{popov86}. It is also widely observed that a peak and
a dip may appear close to each other forming an asymmetric line
shape, a phenomenon called Fano
resonance~\cite{fano41,hessel,popov86,fan03}. For lossy periodic   
structures, a particular incident wave may induce no reflected and transmitted waves,
giving rise to perfect absorption~\cite{hutley76}. When there are more
than one propagating diffraction orders, it is sometimes possible to
force reflected (or transmitted) wave to a particular diffraction
order, leading to the so-called blazing
diffraction phenomenon~\cite{popov01}. These and other special diffraction 
conditions are often referred to as diffraction anomalies, and they have been 
extensively investigated both theoretically and experimentally~\cite{krasnok19}.  

To investigate their properties and realize their applications,
efficient numerical methods are needed to 
calculate diffraction anomalies. The diffraction of a
time-harmonic incident wave by a given periodic structure  is a
boundary value problem (BVP).
Many numerical methods have been developed to solve this
BVP~\cite{li97,bao05,yuexia06,bao10,shen14,wangtao14}. Its general
solution  can be represented by a scattering matrix. As we shall show in
Sec.~II, every diffraction anomaly is equivalent to 
a condition on one
or more entries of the scattering matrix. Therefore, the diffraction
anomalies may be found by searching the frequency (or wavevector, or other 
parameters) iteratively. In each iteration, the BVP is solved to find 
some entries of the scattering matrix. Such an iterative scheme is
widely used, but it may fail if good initial guesses are not 
available. In addition, if all diffraction anomalies (of certain type)
in a given frequency range are required, it is necessary
to  densely scan the frequency interval. The problem becomes more complicated if the
diffraction anomaly only exists when some structural parameters are
properly tuned. Clearly, the existing 
method for computing  diffraction anomalies is not very reliable and
computationally expensive. 

In this paper, we consider a few diffraction anomalies including zero
reflection, zero transmission, perfect absorption and blazing
diffraction, and reformulate all of them as a 
nonlinear eigenvalue problem (NEP)
\begin{equation}
  \label{nep}
  A(\omega) {\bf u} = {\bf 0}
\end{equation}
where $A$ is an operator that can be approximated by a square matrix,
and $\omega$ is the unknown frequency (the eigenvalue) that appears in $A$
nonlinearly. For Eq.~(\ref{nep}) to have a nonzero solution ${\bf u}$,
the operator $A(\omega)$ must be singular. This implies that $\omega$
can be solved from conditions such as $\det A(\omega) = 0$ or
$\sigma_1 (A(\omega)) = 0$, where $\sigma_1$ is the smallest
singular value of $A$. However, such a method also requires
good initial guesses, and has similar disadvantages as the iterative
method based on  the scattering matrix. Our approach is to 
solve the NEP by the contour-integral method developed by
Asakura {\it   et al.}~\cite{asak09} and Beyn~\cite{beyn12}. The
method allows us to find all eigenvalues of
Eq.~(\ref{nep}) inside a closed contour in the complex plane of 
$\omega$. Since no initial guesses are needed, the method is robust and
reliable. 

The rest of this paper is organized as follows. In Sec.~II, we
introduce the diffraction problem and scattering matrix, and identify
the diffraction anomalies as special conditions on one
or more entries of the scattering matrix.  In Sec.~III,  the diffraction anomalies are
reformulated as NEPs. Numerical examples for periodic arrays of
circular cylinders are presented in Sec.~IV. The paper is concluded with a few remarks in
Sec.~V.

\section{Diffraction anomalies}

In a two-dimensional (2D) structure that is translationally invariant
in spatial variable $x$, a polarized electromagnetic wave is also invariant in
$x$, and has only one nonzero component, namely the $x$ component, in 
its electric or magnetic field. We consider an $E$-polarized
time-harmonic electromagnetic wave in a 2D structure with a single
periodic direction. The structure is periodic in $y$, finite in $z$,
and surrounded by air.  The dielectric function $\epsilon$ of the structure 
satisfies 
\begin{eqnarray}
  &&  \label{periodic}
     \epsilon(y+L,z) = \epsilon(y,z), \quad \forall (y,z) \in
     \mathbb{R} \\
  \label{air}
  &&  \epsilon(y,z) = 1, \quad |z| > d, 
\end{eqnarray}
where $L$ is the period in the $y$ direction and $2d$ is the thickness
of the structure. The $x$ component of the electric field, denoted as
$u$, satisfies
\begin{equation}
  \label{helm}
  \partial_y^2 u + \partial_z^2 u + k_0^2 \epsilon(y,z)\, u = 0,
\end{equation}
where $k_0=\omega/c$ is the free space wavenumber, $\omega$ is the
angular frequency, $c$ is the speed of light in vacuum, and the time
dependence is $e^{-i \omega t}$. 

A diffraction problem can have one or more plane incident waves illuminating on
the periodic structure. We assume the incident waves are associated
with a fixed frequency $\omega$ and a fundamental wavenumber $\beta \in
(-\pi/L, \pi/L]$ for the $y$ direction. Due to the periodicity, for
any integer $m$, $\beta_m = \beta + 2\pi m/L$ is a compatible
wavenumber, and the associated plane wave is the $m$-th diffraction
order. For each integer $m$, $\gamma_m = \sqrt{ k_0^2 - \beta_m^2}$ is
either real (non-negative) or pure imaginary. We assume there is a set
of integers $\mathbb{M}$ containing $0$, such that if $m \in \mathbb{M}$ then $\gamma_m
> 0$, if $ m \notin \mathbb{M}$ then $\gamma_m = i \mu_m$ for $\mu_m >
0$.  For $m \in \mathbb{M}$ and $m \notin \mathbb{M}$,
the plane waves with wavevectors $(\beta_m, \pm \gamma_m)$ are propagating
and evanescent diffraction orders, respectively. The diffraction
problem can be studied with incident 
waves containing all propagating diffraction orders and given 
above and below the periodic layer (i.e. for $z> d$ and $z< -d$,
respectively).  For
$|z| \ge  d$, we can write down the solution of a diffraction problem as
\begin{equation}
  \label{Rayleigh}
  u(y,z) = u_{\rm inc}^\pm(y,z) + u_{\rm sca}^{\pm}(y,z), \quad \pm z
  \ge d, 
\end{equation}
where the superscripts  ``$+$'' and ``$-$'' signify waves  above and below
the periodic layer, respectively, and the subscripts indicate incident
and scattered waves, respectively. Moreover, the incident and
scattered  waves can be
expanded in plane waves as
\begin{eqnarray}
&&  \label{inc}
  u_{\rm inc}^{\pm} (y,z) = \sum_{m \in \mathbb{M}} a_m^\pm e^{ i (\beta_m y 
   \mp \gamma_m z)}, \\
  \label{sca}
  && u_{\rm sca}^{\pm} (y,z) = \sum_{m \in \mathbb{Z}}  b_m^\pm e^{ i (\beta_m y 
  \pm \gamma_m z)}, 
\end{eqnarray}
where $\mathbb{Z}$ is the set of all integers, $a_m^\pm$ for $m \in \mathbb{M}$, are given coefficients of the incident
plane waves, and $b_m^\pm$ for all $m$, are the coefficients of the
outgoing propagating or evanescent plane waves.

The scattering matrix $S$ maps the coefficients of the incident waves
to the coefficients of the outgoing propagating waves, namely
\begin{equation}
  \label{Smatrix}
  \begin{bmatrix}
    {\bf b}^+ \cr {\bf b}^-
  \end{bmatrix}
  = S 
    \begin{bmatrix}
    {\bf a}^+ \cr {\bf a}^-
  \end{bmatrix}, 
\end{equation}
where
${\bf a}^+$ is a column vector of $a_m^+$ for all $m \in \mathbb{M}$,
${\bf b}^+$ is a column vector of $b_m^+$ for all $m \in \mathbb{M}$,
etc.  Let $M$ be the number of integers in $\mathbb{M}$, then $S$
is a $(2M) \times (2M)$ square matrix. It is clear that $S$ depends
on the frequency $\omega$ and the wavenumber $\beta$. If the periodic structure has
no material loss, $\epsilon(y,z)$ is a real positive function, then
energy is conserved and $S$ is a unitary matrix. 

The simplest and most important case is $\mathbb{M} = \{ 0 \}$,
namely, the zeroth diffraction order is the only propagating order. In 
that case, $S$ is a  $2 \times 2$ matrix satisfying
\begin{equation}
  \label{S2by2}
  \begin{bmatrix}
    b_0^+ \cr b_0^-
  \end{bmatrix}
  = S 
    \begin{bmatrix}
      a_0^+ \cr a_0^-
  \end{bmatrix}.
\end{equation}
Let $s_{jk}$ be the $(j,k)$ entry of above $S$. If for a fixed
$\beta$, $s_{11}(\omega) =0$ for a real frequency $\omega$, then
for an incident wave given above the periodic layer, there is no
reflected wave.  This is the simplest case of zero reflection and it is considered as a diffraction
anomaly. If the structure is lossless, the unitarity of $S$ implies
$|s_{21}(\omega)|=1$, thus, zero reflection implies total
transmission. Similarly, if $s_{21}(\omega)=0$, then 
$\omega$ is the frequency for zero transmission for an incidence wave
given above the periodic layer. If the 
periodic structure is lossless, zero transmission implies total
reflection. We also regard zero transmission as a diffraction
anomaly. Popov {\it et al.}~\cite{popov86} first realized that structural
symmetry is important to the appearance of zero reflection and zero
transmission. In some cases, the frequency for these anomalies
can be approximated~\cite{popov86,fan03,blan16,wu22}. However, even for periodic
structures with the right symmetry, the existence of zero
reflection/transmission has only been rigorously established for special
circumstances~\cite{shipman12,zero22}.

For lossless structures, the power of incident waves is completely 
converted to outgoing waves. For $M=1$, that means $|a_0^+|^2 + |a_0^-|^2 =
|b_0^+|^2 + |b_0^-|^2$. If the structure has material loss, i.e.,
$\mbox{Im}(\epsilon)$ is positive somewhere, there could be a real
frequency $\omega$ such that $b_0^+ = b_0^- = 0$ 
for some $(a_0^+, a_0^-) \ne (0, 0)$.  This is a case of perfect
absorption~\cite{hutley76,popov08,landy08}  or coherent perfect absorption~\cite{chong10}, and it is a
useful diffraction anomaly for solar cell technology. 
Notice that perfect absorption corresponds to $S(\omega)$ being a
singular matrix, and thus $\det 
S(\omega) = 0$.

If there are two propagating diffraction orders, i.e., $M=2$, then $S$ is a $4
\times 4$ matrix. For $\beta \in (0, \pi/L)$, we have $\mathbb{M} = \{
0, -1 \}$, namely, the  propagating orders correspond to wavenumbers
$\beta_0 = \beta$ and $\beta_{-1}=\beta-2\pi/L$. We are interested in
a diffraction anomaly where incident waves are given in the  
zeroth diffraction order and outgoing waves appear in the 
$-1$st diffraction order only. To write down a condition for this 
{\it blazing} diffraction phenomenon~\cite{popov01}, we assume
vectors ${\bf a}^\pm$ and ${\bf 
  b}^\pm$ are
\begin{equation}
  \label{m2case}
  {\bf a}^{\pm} =
  \begin{bmatrix}
    a_{-1}^{\pm} \cr
    a_0^{\pm} 
  \end{bmatrix}, \quad
  {\bf b}^{\pm} =
  \begin{bmatrix}
    b_{-1}^{\pm} \cr
    b_0^{\pm} 
  \end{bmatrix}, 
\end{equation}
and the entries of $S$ are $s_{jk}$ for 
$1 \le j, k \le 4$. The scattering matrix satisfies 
\begin{equation}
  \label{blaze1}
  \begin{bmatrix}
    b_{-1}^+ \cr 0 \cr b_{-1}^{-} \cr 0
  \end{bmatrix}
= S 
\begin{bmatrix}
  0 \cr a_0^+ \cr 0 \cr a_0^{-}
\end{bmatrix}.
\end{equation}
The 2nd and 4th rows of above give 
\begin{equation}
  \label{blaze2}
\begin{bmatrix}
  s_{22} & s_{24} \cr s_{42} & s_{44}
\end{bmatrix}  
\begin{bmatrix}
  a_0^+ \cr a_0^{-}
\end{bmatrix}
=
\begin{bmatrix}
  0 \cr 0
\end{bmatrix}.
\end{equation}
Therefore,
\begin{equation}
  \label{blaze3}
  \det \begin{bmatrix}
  s_{22} & s_{24} \cr s_{42} & s_{44}
\end{bmatrix}  
= 0.
\end{equation}

In summary, we have considered diffraction anomalies including zero
reflection, zero transmission, perfect absorption and blazing
diffraction.  In terms of the scattering matrix, each anomaly corresponds
to a zero condition on an entry or a sub-matrix of the scattering
matrix.  The diffraction anomalies may occur at a
specific frequency $\omega$ for a fixed structure and a fixed wavenumber
$\beta$. In that case, all we have to do is to calculate $\omega$ by
solving a scalar equation. If the incident angle is fixed, then
$\beta$ is related to $\omega$, and the frequency is still the only unknown.
Diffraction anomalies can also be studied for a fixed frequency, then the
wavenumber $\beta$ or incident angle is the unknown. However, for a
fixed structure, some diffraction anomalies 
may not occur for any frequency or wavenumber. In that case,
it is necessary to add tunable parameters to the structure, and solve
the parameters together with the frequency and/or wavenumber. 

In principle, we can find  diffraction anomalies by solving
the equations obtained from the scattering matrix. Such a method works
well if  there are good initial guesses, but since good initial
guesses are not easy to obtain, the method is not robust. Moreover,  the method
becomes computationally expensive if all diffraction anomalies in
a given frequency or wavenumber interval  are required. 
To obtain good initial guesses, it is necessary to 
densely scan the frequency or wavenumber interval. This implies that the diffraction
problem must be solved repeatedly for many different values of the
frequency or wavenumber, and this is computationally expensive. 

\section{Nonlinear eigenvalue formulations}

To overcome the difficulty of finding good initial guesses for all
diffraction anomalies in a given frequency/wavenumber  interval, we develop a
robust numerical method  based on
nonlinear eigenvalue formulations and a contour-integral
method for solving nonlinear eigenvalue
problems~\cite{asak09,beyn12}. Although the diffraction of a
time-harmonic wave is a BVP, the diffraction anomalies are special
conditions of this BVP, and they can be reformulated as eigenvalue
problems where the eigenvalue is the frequency or wavenumber. However, the
eigenvalue problem is nonlinear with a nonlinearity in the
eigenvalue. Fortunately, this type of nonlinear eigenvalue problems
(NEPs) can be accurately and robustly solved by the contour-integral
method~\cite{asak09,beyn12}. 

To describe the NEP formulations for various diffraction anomalies, we
first consider the standard eigenvalue problem for resonant
modes~\cite{fan02,link19}, and reformulate this linear eigenvalue
problem as a NEP. In a 2D 
periodic structure given by a  dielectric function $\epsilon(y,z)$ satisfying conditions
(\ref{periodic}) and (\ref{air}), any $E$-polarized eigenmode is a
Bloch mode
\begin{equation}
  \label{bloch}
  u(y,z) = \phi(y,z) e^{i \beta y}, 
\end{equation}
where $\phi$ is periodic in $y$ with period $L$ and $\beta$ is the
Bloch wavenumber. The eigenvalue
problem is for $u$ satisfying Eq.~(\ref{helm}) and proper boundary
conditions as $z \to \pm \infty$.  The eigenvalue is either $\omega$
(or $k_0=\omega/c$, or $k_0^2$) for given real $\beta \in (-\pi/L,
\pi/L]$, or $\beta$ for given $\omega > 0$. 
We focus on the case
where $\omega$ is the eigenvalue. If the boundary condition is $u \to
0$ as $z \to \pm \infty$, then the eigenmode is a guided mode. We are
concerned with resonant modes (also called resonant states or
quasi-normal modes) for which $u$ satisfies an outgoing radiation condition
as $z \to \pm \infty$~\cite{fan02,link19}. This condition implies that
power is radiated out to infinity as $z \to \pm \infty$, and
$\omega$ must have a negative imaginary part, so that the mode
amplitude decays with time.  

For $|z| > d$, the nonzero electric field component of a resonant mode
can be expanded in plane waves, exactly like $u_{\rm sca}^\pm$ in
Eq.~(\ref{sca}), namely, 
\begin{equation}
  \label{expand}
  u(y,z) = \sum_{m\in \mathbb{Z}} c_m^\pm e^{ i (\beta_m y  \pm 
    \gamma_m z)}, \quad \pm z \ge d, 
\end{equation}
where $c_m^\pm$ are the expansion coefficients, $\beta_m$ and
$\gamma_m$ are given in Sec.~II. However, $k_0$ is now complex with a
negative imaginary part, $k_0^2 - \beta_m^2$ is in the lower half of the
complex plane,  the complex square root in 
\begin{equation}
  \label{gamma}
  \gamma_m = \sqrt{ k_0^2 - \beta_m^2}
\end{equation}
should be defined using a branch cut along the negative imaginary axis (instead of
the negative real axis), so that
when  $k_0^2 - \beta_m^2$ is in the third or fourth quadrant, 
$\gamma_m$ is in the second or fourth quadrant, respectively. 
This choice of complex square root ensures that each term in the right hand side of
Eq.~(\ref{expand}) is either
an evanescent plane wave that 
decays exponentially as $z \to \pm \infty$, or 
an outgoing plane wave that radiates out power (and grows 
exponentially) as $z \to \pm \infty$. 
As in Sec.~II, we have a
set $\mathbb{M}$ for those integers $m$ such that $k_0^2 - \beta_m^2$ is in
the fourth quadrant. We assume $\mathbb{M}$ is not empty and contains
$0$. In that case, the resonant mode has the following  far field
asymptotic expansion 
\begin{equation}
  \label{farfield}
  u(y,z) \sim \sum_{m \in \mathbb{M} } c_m^\pm e^{ i (\beta_m y  \pm 
    \gamma_m z)}, \quad   z \to \pm \infty. 
\end{equation}
The linear eigenvalue problem of a resonant mode is for $u$ satisfying
Eq.~(\ref{helm}) in $\Omega_{\rm inf}$,
the far field condition (\ref{farfield}), and the following
quasi-periodic conditions
\begin{equation}
  \label{qperiod}
  \begin{bmatrix}
    u \cr 
\partial_y u
  \end{bmatrix}_{z=L}
  = e^{ i \beta L}
  \begin{bmatrix}
    u \cr 
    \partial_y u
  \end{bmatrix}_{z=0}, 
\end{equation}
where $\Omega_{\rm inf}$ is given by $0<y<L$ and $-\infty < z < \infty$.
Since the coefficients $c_m^\pm$ for $m \in \mathbb{M}$ are unknown, the far field
condition (\ref{farfield}) is difficult to use. The standard approach
is to use the perfectly matched layer (PML) technique~\cite{pml94,chew94,pml97}, namely, 
move $z$ to a path in the complex plane so that $u(y,z) \to 0$ as 
$z \to \infty$ along the path.

If we define a linear operator $\Lambda_0$ such that 
\begin{equation}
  \label{lam0}
  \Lambda_0 e^{ i \beta_m y} = i \gamma_m e^{ i \beta_m y}, \quad m
  \in \mathbb{Z}, 
\end{equation}
then Eq.~(\ref{expand}) gives rise to 
\begin{eqnarray}
&&  \label{bcd}
  \frac{ \partial u}{\partial z}   = \Lambda_0 u, \quad  z= d, \\
&&  \label{bcnd}
  \frac{ \partial u}{\partial z}   = -\Lambda_0 u, \quad  z= -d.
\end{eqnarray}
The operator $\Lambda_0$ maps $u$ (Dirichlet data) to the derivative
of $u$ (Neumann data), and is a so-called Dirichlet-to-Neumann (DtN)
operator.  Since $\Lambda_0$ depends on $\omega$ and $\beta$, we obtain a NEP
for $u$ satisfying Eq.~(\ref{helm}) in $\Omega_d$, and boundary
conditions (\ref{qperiod}), (\ref{bcd}) and (\ref{bcnd}), where $\Omega_d$ is the
rectangular domain given by $0 < y < L$ and $-d < z < d$.

The above NEP formulation on $\Omega_d$ can be used for numerical
implementation, but we prefer a NEP formulated on two line segments at $z= \pm d$ (for $0 < y < L$). To achieve this, we define a
linear operator $F$ that maps $u$ at $z=\pm d$ (as functions of $y$ 
for $0 < y < L$) to  $\partial_z u$ at $z=\pm d$, where $u$
satisfies Eq.~(\ref{helm}) and boundary condition (\ref{qperiod}).
The operator $F$ depends on both $\beta$ and $\omega$,  and satisfies 
\begin{equation}
  \label{defF}
  F  {\bf u} 
  =
  \begin{bmatrix}
    \partial_z u (y,d)  \cr  \partial_z u (y,-d)
  \end{bmatrix},
  \quad
  {\bf u} =
  \begin{bmatrix}
    u(y,d) \cr u(y,-d) 
  \end{bmatrix}. 
\end{equation}
Let $F$ be given
in $2 \times 2$ blocks,
\begin{equation}
  \label{F2by2}
  F =
  \begin{bmatrix}
    F_{11} & F_{12} \cr F_{21} & F_{22}
  \end{bmatrix}, 
\end{equation}
then Eqs.~(\ref{bcd})-(\ref{defF}) lead to Eq.~(\ref{nep}), 
where $A=A(\omega)$ is the $2\times 2$ matrix operator
\begin{equation}
  \label{matA0}
  A =
  \begin{bmatrix}
    F_{11}-\Lambda_0 & F_{12} \cr F_{21} & F_{22} + \Lambda_0 
  \end{bmatrix}. 
\end{equation}
Equation~(\ref{nep}) with the above $A$ is our preferred NEP formulation for resonant
modes. For given $\omega$ and $\beta$, the two operators $\Lambda_0$
and $F$ can be approximated by matrices. If $ y \in (0,L)$ is discretized by $N$
points, then ${\bf u}$ is approximated by a column vector of length
$2N$, $\Lambda_0$ and $F$ are approximated by $N \times N$ 
and $(2N) \times (2N)$ matrices, respectively. In Appendix, we give
additional details on computing the matrix approximations of
$\Lambda_0$ and $F$. Since we assume $\beta$
is given and $\omega$ is the unknown, we emphasize the dependence on
$\omega$ by writing $A$ as $A(\omega)$ in Eq.~(\ref{nep}). 

Next, we present NEP formulations for diffraction anomalies
discussed in Sec.~II. First, we
consider a zero reflection where the incident wave is given 
above the periodic layer (i.e., for $z > d$) in the zeroth diffraction
order and there is no reflected wave in the same diffraction
order. Therefore, the total field for $|z|>d$ can be expanded as follows:
\begin{eqnarray}
  \label{zeroR0}
&&  u  = a_0^+ e^{i (\beta_0 y - \gamma_0 z)}
   + \sum_{m\ne 0} b_m^+ e^{i (\beta_m y + \gamma_m z)}, \ z > d,
   \quad \\  
&&  u = \sum_{m \in \mathbb{Z}}  b_m^- e^{i (\beta_m y -
   \gamma_m z)}, \ z < -d. 
\end{eqnarray}
Clearly,  $u$ satisfies Eq.~(\ref{bcnd}), the same boundary
condition as the resonant modes,  at $z=-d$. To obtain a boundary condition at
$z=d$, we define a new linear operator $\Lambda_1$ by 
\begin{equation}
  \label{lam1}
  \Lambda_1 e^{ i \beta_m y}
  =
  \begin{cases}
    -i \gamma_0 e^{i \beta_0 y},   \quad \text{$m=0$}, \\
    i \gamma_m e^{i \beta_m y}, \quad   \text{ $m\ne 0$}, 
  \end{cases}
\end{equation}
then $u$ satisfies
\begin{equation}
  \label{bcd1}
  \frac{\partial u}{\partial z} = \Lambda_1 u, \quad z=d. 
\end{equation}
Therefore, the NEP  for zero reflection is Eq.~(\ref{nep}) with a new
matrix operator $A$ given by
\begin{equation}
  \label{matA1}
  A =
  \begin{bmatrix}
    F_{11}-\Lambda_1 & F_{12} \cr F_{21} & F_{22} + \Lambda_0 
  \end{bmatrix}.
\end{equation}

The case of zero transmission is somewhat more complicated. If
an incident wave is given above the periodic layer in the zeroth
diffraction order, and there is no transmitted wave in the zeroth diffraction order
below the layer, then the total field is
\begin{eqnarray}
  \label{zeroT0}
&&  u  = a_0^+ e^{i (\beta_0 y - \gamma_0 z)}
   + \sum_{m\in \mathbb{Z}}  b_m^+ e^{i (\beta_m y + \gamma_m z)}, \ z
   > d, \quad  \\
&&  u = \sum_{m\ne 0} b_m^- e^{i (\beta_m y -
   \gamma_m z)}, \ z < -d.
\end{eqnarray}
Because of the incident wave, the boundary condition at $z=d$ is
inhomogeneous.
We have
\begin{equation}
  \label{bct0}
\frac{\partial u}{\partial z} = \Lambda_0 u - 2 i a_0^+ \gamma_0 e^{ i
  (\beta_0 y - \gamma_0 d)}, \quad z=d.
\end{equation}
Although $b_0^-=0$, $u$ still satisfies Eq.~(\ref{bcnd}) at $z=-d$. The condition $b_0^-=0$ implies that 
\begin{equation}
  \label{zeroT0cond}
  \int_0^L u(y,-d) e^{- i \beta_0
     y}\, dy = 0. 
\end{equation}
Combining the above with the operator $F$, we obtain a NEP given as
Eq.~(\ref{nep}) with a new matrix operator
\begin{equation}
  \label{nepT0}
A=   \begin{bmatrix}
    F_{11} - \Lambda_0 & F_{12}  & f(y) \cr 
    F_{21} & F_{22} + \Lambda_0  & 0 \cr 
    0 & {\sf g} & 0 
  \end{bmatrix},
\end{equation}
and a new vector 
\begin{equation}
{\bf u} =   
  \begin{bmatrix}
    u(y,d) \cr
    u(y,-d) \cr
    1
  \end{bmatrix},
  \label{uwith1}
\end{equation}
where $f(y) = 2i a_0^+ \gamma_0 e^{i (\beta_0 y - \gamma_0 d)}$,  and
${\sf g}$ is the linear functional that maps $u(y,-d)$ to the left
hand side of  Eq.~(\ref{zeroT0cond}). When $y \in (0, L)$ is
discretized by $N$ points, $f(y)$ becomes  by a column vector
of length $N$, ${\sf g}$ is approximated by a row vector of length
$N$, and $A$ becomes a $ (2N+1) \times (2N+1)$ matrix. 

When the structure is absorptive, we can consider perfect absorption
for which some incident waves do not produce any outgoing propagating 
waves. If there is only one propagating diffraction order for each
side of the periodic layer, the total field can be written as
\begin{eqnarray}
  \label{pabs1}
&&  u  = a_0^+ e^{i (\beta_0 y - \gamma_0 z)}
   + \sum_{m\ne 0} b_m^+ e^{i (\beta_m y + \gamma_m z)}, \ z > d, \\
  \label{pabs2}
  &&  u = a_0^- e^{i (\beta_0 y + \gamma_0 z)}
     + \sum_{m\ne 0}  b_m^- e^{i (\beta_m y -
   \gamma_m z)}, \ z < -d. \quad
\end{eqnarray}  
Therefore,  the boundary condition at $z=d$ is Eq.~(\ref{bcd1}), 
same as the zero reflection case considered above. The boundary condition at
$z=-d$ is 
\begin{equation}
  \label{pabsbc}
  \frac{\partial u}{\partial z} =  - \Lambda_1 u, \quad z=-d. 
\end{equation}
Therefore, the NEP for perfect absorption is Eq.~(\ref{nep}) with ${\bf u}$ given in
(\ref{defF}), and $A$ given by
\begin{equation}
  \label{Apabs}
  A =
  \begin{bmatrix}
    F_{11}-\Lambda_1 & F_{12} \cr F_{21} & F_{22} + \Lambda_1
  \end{bmatrix}. 
\end{equation}

For a lossless periodic structure, and if there are two propagating
diffraction orders (the zeroth  and negative-first orders if $0 < \beta
\le \pi/L$), we can consider a blazing diffraction phenomenon that converts all
power of the incident waves (in zeroth diffraction order) to outgoing
waves in the negative first order. The total field for $|z| > d$ has the
same expansions (\ref{pabs1}) and (\ref{pabs2}). Therefore, the
NEP formulation is also Eq.~(\ref{nep}) with $A$ given in
Eq.~(\ref{Apabs}) and ${\bf u}$ given in (\ref{defF}).

\section{Numerical examples}

To illustrate our method, we consider a periodic array of dielectric
cylinders surrounded by air. The radius and the dielectric constant of
the cylinders are $a$ and $\epsilon_1$, respectively.  The cylinders
are parallel to the $x$ axis. Their centers are located on the $y$ axis.
The period $L$ of the array is the distance between the centers 
of two nearby cylinders. The array is considered as a periodic layer
with a thickness $2d = L$. For all numerical examples, we discretize
the interval $(0, L)$ by $N=11$ points. Therefore,  the
operators $\Lambda_0$ and $\Lambda_1$ are approximated by $11 \times 11$
matrices, and the operator $F$ is approximated by a $22 \times 22$
matrix, and the matrix $A$ is either $22 \times 22$ or $23 \times
23$. When the contour-integral method is used to solve the NEP, we use
100 points to discretize the contour and approximate integrals along
the contour by the trapezoid method. 

\begin{figure}[h]
  \centering
  \includegraphics[width=\linewidth]{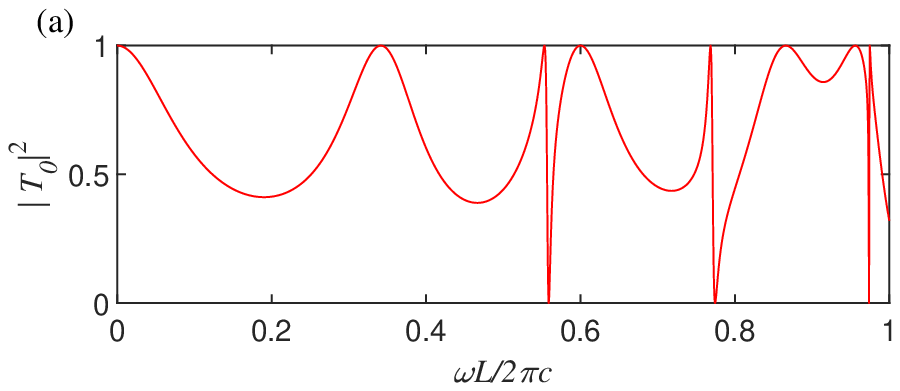}
  \includegraphics[width=\linewidth]{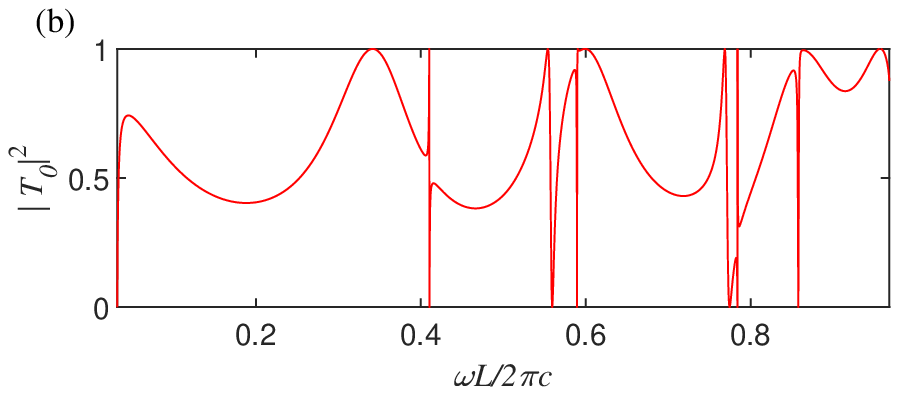}
  \caption{Transmission spectra of a periodic array of circular
    cylinder with radius $a=0.3L$ and dielectric constant
    $\epsilon_1=11.6$ for, (a) a normal incident wave, (b) an incident plane
    wave with $\beta = 0.2/L$.}
  \label{fig:Tspectra}
\end{figure}

First, we consider the periodic array with $a = 0.3L$ and
$\epsilon_1=11.6$. In Fig.~\ref{fig:Tspectra},
we show the transmission spectra for a plane incident wave with
wavenumber $\beta=0$ (normal incidence) and 
$\beta = 0.2/L$, respectively.  The incident wave is given above the 
array, thus $a_0^+\ne 0$, $a_0^{-}=0$ and the transmission
coefficient is $T_0 = s_{21} = b_0^{-}/a_0^{+}$. From
Fig.~\ref{fig:Tspectra}, it appears that total transmission (zero reflection) and zero
transmission occur at some special frequencies.
To find the frequencies for zero reflection/transmission and the
resonant modes, we use the contour-integral method with a circular
contour in the complex plane of $\omega$ centered on the real axis. 
In Table~\ref{tab:res}, 
\begin{table}[h]
  \caption{Frequencies of resonant modes 
    in  a periodic array of circular cylinders with $a=0.3L$ and
    $\epsilon_1 = 11.6$, for  wavenumber $\beta=0$ and $ 0.2/L$.} 
  \centering 
  \begin{tabular}{|c||c|} \hline 
    $\omega L/(2\pi c)$ for $\beta = 0$  &   $\omega L/(2\pi c)$ for $\beta = 0.2/L$  \\ \hline 
\ $0.557333  - 0.002647i$ \ & \ $0.557898 - 0.002502i$ \ \\ \hline 
${\it 0.589733}$ & $0.589439 - 0.000147i$\\ \hline 
$0.593629 - 0.058809i$ & $0.593758 - 0.058677i$\\ \hline 
$0.770917 - 0.002940i$ & $0.771034 - 0.002876i$\\ \hline 
${\it 0.784154}$ & $0.784059 - 0.000061i$\\ \hline 
${\it 0.858999}$ & $0.857857 - 0.000543i$\\	\hline 
$0.857873 - 0.054473i$ & $0.857930 - 0.054206 $\\ \hline 
  \end{tabular}
  \label{tab:res}
\end{table}
we show seven resonant modes for $\beta =0$ and $\beta
= 0.2/L$, respectively. The results are obtained using a contour with
center at $0.7 (2\pi c/L)$ and radius $0.2 (2\pi c/L)$. 
Notice that for $\beta=0$, there are three
special resonant modes with a real frequency. They are examples of bound states in the  
continuum
(BICs)~\cite{hsu16,sad21,shipman03,port05,mari08,hsu13,bulg14,hu15}.
A BIC has the same expansion, Eq.~(\ref{expand}), as the resonant modes, but
$c_m^\pm = 0$ for any $m$ such that $\gamma_m$ is real. For the three BICs in
Table~\ref{tab:res}, since $\mathbb{M} = \{ 0 \}$, only $c_0^\pm =
0$. Although a BIC does not radiate out power as $z \to \pm \infty$,
it satisfies the same boundary conditions and the same NEP 
formulation as the resonant modes. The three BICs in
Table~\ref{tab:res} are standing waves with $\beta=0$, and they turn
to resonant modes with a high $Q$ factor as $\beta$ moves away from
zero. 

In Table~\ref{tab:r0},
\begin{table}[h]
  \caption{Frequencies of zero reflection in a periodic array circular 
    cylinders with $a=0.3L$ and  
    $\epsilon_1 = 11.6$, for wavenumber $\beta=0$ and $0.2/L$.}
  \centering 
  \begin{tabular}{|c||c|} \hline 
    $\omega L/(2\pi c)$ for $\beta = 0$  & $\omega L/(2\pi c)$ for $\beta = 0.2/L$  \\ \hline 
$0.553305$ & $0.554087$\\ \hline 
    $0.600099$    & $0.590294$\\	\hline 
    $0.768452$     & $0.599160$\\ \hline 
    $0.865895$     & $0.768661$\\	\hline 
               --        & $0.784094$\\ \hline 
  \end{tabular}
  \label{tab:r0}
\end{table}
we list zero-reflection frequencies for both $\beta=0$ and 
$\beta=0.2/L$.
In Fig.~\ref{fig:Tspectra}, we can find seven and nine total-transmission
frequencies in $(0, 2\pi c/L)$ for $\beta=0$ and $\beta=0.2/L$,
respectively. In Table~\ref{tab:r0}, only four and five  zero-reflection
frequencies are listed, since they are the ones inside the contour
chosen for the computation. For $\beta = 0$, we use the
same circular contour as before [with center at $0.7 (2\pi c/L)$ and radius $0.2 (2\pi
c/L)$], and obtain all four zero-reflection frequencies in the
interval $(0.5, 0.9)(2\pi c/L)$, as well as three BICs listed in
Table~\ref{tab:res}. Interestingly, the BICs also satisfy the boundary
conditions for the zero reflection solutions. For $\beta=0.2/L$, the
chosen contour is centered at $0.65 (2\pi c/L)$ 
and has a radius $0.15 (2\pi c/L)$, thus the obtained zero-reflection
frequencies belong to the interval $(0.5, 0.8) (2\pi c/L)$.

From Fig.~\ref{fig:Tspectra}, it appears that there are three and six 
zero-transmission frequencies in the interval $(0, 2\pi c/L)$ for $\beta=0$
and $\beta=0.2/L$, respectively. To find all zero-transmission
frequencies in $(0.3, 0.9) (2\pi c/L)$, we use a circular contour with
center at $0.6 (2\pi c/L)$ and radius $0.3 (2\pi c/L)$. The results
are listed in Table~\ref{tab:t0}
\begin{table}[h]
  \caption{Frequencies of zero transmission in a periodic array circular 
    cylinders with $a=0.3L$ and  $\epsilon_1 = 11.6$, for wavenumber
    $\beta=0$ and $0.2/L$.} 
  \centering 
  \begin{tabular}{|c||c|} \hline 
    $\omega L/(2\pi c)$ for $\beta = 0$  & $\omega L/(2\pi c)$ for $\beta = 0.2/L$  \\ \hline 
    $0.558859$ &   $0.410497$ \\ \hline
    $0.774310$ & $0.559341$ \\ \hline
- & $0.589390$ \\ \hline 
-& $0.774413$  \\ \hline 
- & $0.783940$  \\ \hline 
- & $0.857738$  \\ \hline    
  \end{tabular}
  \label{tab:t0}
\end{table}
for both $\beta=0$ and $\beta=0.2/L$.
The NEP for zero transmission, i.e. Eq.~(\ref{nep}) for $A$ given in
(\ref{nepT0}), is also satisfied by the BICs, but unlike that given in
Eq.~(\ref{uwith1}), for a BIC, the last entry of ${\bf u}$ is 
zero. For $\beta=0$, the numerical results include two
zero-transmission frequencies listed in Table~\ref{tab:t0} and four
BICs [three listed in Table~\ref{tab:res} and a new one with
frequency $\omega = 0.411228 (2\pi c/L)$]. Of course, the computed vector ${\bf
  u}$ is scaled differently, its last entry is not  simply 1 or
0. However, the BICs can be easily identified by considering the ratio
between the last entry and the entry with the 
maximum magnitude. In our case, the ratio for the four BICs ranges from ${\cal
  O}(10^{-9})$ to ${\cal O}(10^{-7})$. The contour-integral method
generally gives complex solutions for $\omega$. For the two
zero-transmission frequencies of $\beta=0$ in Table~\ref{tab:t0}, the
imaginary part of the normalized frequency $\omega L/(2\pi c)$ is
${\cal O}(10^{-11})$. For $\beta=0.2/L$, the numerical solutions are
less accurate, but $\mbox{Im}(\omega) L/(2\pi c)$ is still bounded by
$4.4\times 10^{-5}$ for all cases listed in the right column of
Table~\ref{tab:t0}.  

Next, we consider perfect absorption of normal incident waves ($\beta=0$) in a periodic array of circular
cylinders with material loss, where
$\epsilon_1$  (the dielectric constant of the cylinders) is 
complex. However, for a fixed $\epsilon_1$ and a fixed radius $a$,
perfect absorption does not usually occur, and the NEP problem for $A$
given in Eq.~(\ref{Apabs}), has only complex-$\omega$ solutions.  To find perfect absorption for the
periodic array, we have to tune a structural or material
parameter. For example, if $a=0.3L$ is fixed and the refractive index
of the cylinders is $n_1 = \sqrt{\epsilon_1} = \sqrt{11.6} + i
\sigma$, where $\sigma$ is a parameter, then perfect absorption
occurs at $\sigma= 0.0142765$ with a real frequency
$\omega =0.770981 (2\pi c/L)$.
This result is obtained iteratively
with an iteration in $\sigma$. In each iteration (i.e. for a given
$\sigma$), we solve the NEP using a circular
contour (in the complex $\omega$ plane) with center $0.8 (2\pi c/L)$
and radius $0.1 (2\pi c/L)$ and find a solution $\omega$ which is
complex in general. The iterative process can be regarded as a root-finding method for
solving $\mbox{Im}(\omega) = 0$.
We can also find perfect absorption by tuning the radius $a$ for a
fixed complex $\epsilon_1$. In Table~\ref{tab:pa}, 
\begin{table}[h]
  \caption{Radius $a$ and frequency $\omega$ for perfect absorption in
    a periodic array of lossy circular  cylinders with refractive index
    $n_1 = \sqrt{11.6} + i\sigma$.}
  \centering
  \begin{tabular}{|c|c|c|} \hline
  \  $\sigma=\mbox{Im}(n_1)$ \ & $a/L$ & \ $\omega L/(2\pi c)$ \ \\ \hline
    0.012  &  \ 0.322575\   &   0.715869 \\ \hline
    0.013 &  0.312982 &    0.738371 \\ \hline
    0.014 &  0.302888 &   0.763499  \\ \hline
    0.015 &  0.292248 &  0.791757 \\ \hline
  \end{tabular}
  \label{tab:pa}
\end{table}
we list a few cases where the imaginary part of $n_1$ is
specified. The results are obtained using a circular contour with
center $0.7 (2\pi c/L)$ and radius $0.2 (2\pi c/L)$. 
	
Finally, we consider blazing diffraction for a periodic array of
cylinders with radius $a = 0.3L$ and dielectric constant $\epsilon_1 =
15.42$, and concentrate on the case of two propagating diffraction orders with
wavenumbers $\beta=\beta_0 = \pi/L$ and $\beta_{-1} = -\pi/L$. 
Using a circular contour with center $0.65 (2 \pi c/L)$ 
and radius $0.11 (2\pi c/L)$, we find the following five frequencies 
\[
  0.549336, \
  0.670542, \
  0.678285, \
  0.686822, \
  0.745395.
\]
Since the periodic array has a mirror symmetry in $z$, the blazing
diffraction solutions are either even or odd in $z$. Among the five
solutions above, the first three are even in $z$ and the last two are
odd in $z$.

We have repeated some calculation using different values of $N$ [for
discretizing the interval $(0,L)$] and different number of points for
discretizing the contour. Typically, $N=9$ is sufficient to give four
significant digits. When $N$ is increased to $11$, the accuracy is
improved by at least a  factor of two. For discretizing the contour,
100 points is more than enough. Typical numerical results obtained
with 100 and 200 points for the contour have more than 
eight identical digits.

\section{Conclusion}
Diffraction anomalies such as zero reflection, zero transmission,
perfect absorption and blazing diffraction are interesting wave
phenomena with important applications. Existing methods for computing
the diffraction anomalies either scan the frequency (or wavenumber,
or other parameters) densely or determine the frequency iteratively,
and thus, they are computationally expensive and not very reliable.  Our
method based on NEP formulations and a contour-integral method is capable of
finding all diffraction anomalies in a given frequency interval. Since
the NEPs involve small matrices and no initial guesses for the
frequency are needed, our method is efficient and robust. Although the
method is only formulated for $E$-polarized waves in 2D structures
with a single periodic direction, it can be easily generalized to
full-vector waves in 3D structures with two periodic directions. It is
also straightforward to extend the method to other diffraction
anomalies. The method provides a useful tool for analyzing diffraction
anomalies and explore their applications.

\section*{Acknowledgment}
The authors acknowledge support from the Research Grants Council of
Hong Kong Special Administrative Region,  China (Grant No. CityU 11304619). 

\appendix* 
\section{Matrix approximation of $\Lambda_0$ and $F$}
Here, we briefly describe how opetrators $\Lambda_0$ and
$F$ can be approximated by matrices, if the structure is a periodic
array of circular cylinders surrounded by air. Let $L$ be the period
of the array, $a$ and
$n_1$ be the radius and refractive index of the cylinders,
respectively. For $d=L/2$,  $\Omega_d$ is a square given by
$0<y<L$ and $-d < z < d$. We assume a cylinder is contained in
$\Omega_d$ and centered at $(y,z) =(L/2, 0)$.

Let $N=2p+1$ be a positive odd integer and $y_j = (j-0.5)/L$
for $1\le j \le N$. For $z \ge  d$, the field
expansion (\ref{expand}) may be approximated by
\begin{equation}
  \label{exp0}
  u(y,z) \approx \sum_{m=-p}^p c_m^+ e^{i (\beta_m y + \gamma_m z)}, \quad z 
\ge  d.  
\end{equation}
Evaluating the above at $(y,z) =(y_j,d)$, we obtain an $N \times N$
matrix $C_1$, such that
\[
{\bf u}^+ 
  \approx C_1
{\bf c}^+, 
\]
where ${\bf u}^+$ is a column vector of  $u(y_j, d)$ for $1\le j \le N$, and
${\bf c}^+$ is a column vector of $c_m^+$ for $-p \le m \le p$. 
We can also take a derivative with respect to $z$ for the
approximate expansion above and evaluate the result at $(y_j,d)$ for
$1\le j \le N$. This
gives rise to a matrix $C_2$ such that 
\[
  \partial_z {\bf u}^+
  \approx C_2
{\bf c}^+, 
\]
where  $\partial_z {\bf u}^+$ is a vector for $\partial_z u(y_j, d)$. 
The operator $\Lambda_0$ is then approximated by the $N \times N$
matrix $C_2 C_1^{-1}$.

Inside $\Omega_d$, the general solution Eq.~(\ref{helm}) is
\begin{equation}
  \label{exp2}
  u(y,z) = \sum_{m=-\infty}^{+\infty} d_m \phi_m( r) e^{i m \theta}, 
\end{equation}
where $(r, \theta)$ are the polar coordinates satisfying
\[
  y =L/2 + r \cos \theta, \quad z = r \sin \theta,
\]
and $\phi_m$ is a particular solution with a cylindrical wave incident
upon the cylinder. More specifically,
$\phi_m(r) = A_m J_m (k_0 n_1 r)$ for $ r<a$ and
$\phi_m(r) = B_m H_m^{(1)}(k_0 r) + H_m^{(2)}(k_0 r)$ for $r>a$, where $J_m$,
$H_m^{(1)}$, $H_m^{(2)}$ are $m$-th order Bessel and Hankel
functions. The coefficients $A_m$ and $B_m$ can be solved from the
condition that $\phi_m$ and $d\phi_m/dr$ are continuous at $r=a$. Now,
we approximate the expansion by $4N$ terms:
\begin{equation}
  \label{exp3}
  u(y,z) \approx \sum_{m=-2N}^{2N-1} d_m \phi_m( r) e^{i m \theta}. 
\end{equation}
From the above, we can take partial derivatives and find the
approximate expansions for $\partial_y u$ and 
$\partial_z u$. Let $z_k = -L/2 + (k-0.5)/L$ for $k=1$, 2, ...,
$N$. Evaluating $u$ by Eq.~(\ref{exp3}) at $(y_j, d)$, $(y_j, -d)$ for
$1\le j \le N$ and $(0, z_k)$ and $(L, z_k)$ for $1\le k \le N$, we
obtain a $(4N) \times (4N)$ matrix $D_1$ mapping a column vector ${\bf
  d}$ (for $d_m$, $-2N \le m < 2N-1$) to a column vector of length
$4N$ for $u$ at the $4N$ sampling points on the boundary of
$\Omega_d$. Similarly, we can evaluate
$\partial_z u$ at $(y_j, \pm 
d)$ for $1\le j\le N$ and evaluate
$\partial_y u$ at $(0, z_k)$ and $(L, z_k)$
for $1\le k\le N$, and obtain a $(4N) \times (4N)$ matrix $D_2$ that
maps vector ${\bf d}$ to a vector of length $4N$ for the normal
derivative of $u$ at the $4N$ points on the boundary of
$\Omega_d$. Therefore,
$D_2 D_1^{-1}$ is a $(4N) \times (4N)$ matrix mapping $u$ to the
normal derivative of $u$ at the $4N$ points on the boundary of
$\Omega_d$. Finally, we can use the quasi-periodic condition
(\ref{qperiod}) to eliminate $u$ and $\partial_y u$ at $y=0$ and $y=L$. The
final result is a $(2N) \times (2N)$  matrix $F$ satisfying
\begin{equation}
  \begin{bmatrix}
    \partial_z {\bf u}^+ \cr 
    \partial_z {\bf u}^-
  \end{bmatrix}
  = F
  \begin{bmatrix}
     {\bf u}^+ \cr 
     {\bf u}^-
  \end{bmatrix}, 
\end{equation}
where ${\bf u}^-$ and $\partial_z {\bf u}^-$ are column vectors of
$u(y_j, -d)$ and
$\partial_z u(y_j, -d)$ for $1\le j\le N$, respectively.

\end{document}